\documentclass[]{interact}

\usepackage{epstopdf}% To incorporate .eps illustrations using PDFLaTeX, etc.
\usepackage[caption=false]{subfig}% Support for small, `sub' figures and tables

\usepackage{color}

\usepackage[numbers,sort&compress]{natbib}% Citation support using natbib.sty
\bibpunct[, ]{[}{]}{,}{n}{,}{,}% Citation support using natbib.sty
\renewcommand\bibfont{\fontsize{10}{12}\selectfont}% Bibliography support using natbib.sty

\theoremstyle{plain}% Theorem-like structures provided by amsthm.sty

\theoremstyle{definition}

\theoremstyle{remark}

\newcommand{\Figref}[1]{Figure~\ref{#1}}

\newcommand{\Tabref}[1]{Table~\ref{#1}}

\newcommand{\Eqnref}[1]{Equation~\eqref{#1}}
\newcommand{\Eqnrefs}[1]{Equations~\eqref{#1}}

\usepackage{listings}
\begin{document}

% \articletype{ARTICLE TEMPLATE}% Specify the article type or omit as appropriate

\title{Adjoint computations by algorithmic differentiation of a parallel solver for time-dependent PDEs}

\author{
\name{J.~I. Cardesa\textsuperscript{a}\thanks{CONTACT J.~I. Cardesa. Email: jcardesa@imft.fr}, L. Hasco{\"e}t\textsuperscript{b} and C. Airiau\textsuperscript{a}}
\affil{\textsuperscript{a}IMFT, Universit\'e de Toulouse, UMR 5502 CNRS/INPT-UPS, France \\
\textsuperscript{b}INRIA Sophia-Antipolis, France}
}

\maketitle

\begin{abstract}
A computational fluid dynamics code is
differentiated using algorithmic differentiation (AD) in both tangent and adjoint modes.
The two novelties of the present approach are 1)
the adjoint code is obtained by letting the AD tool Tapenade invert the 
complete layer of message passing interface (MPI) communications, and 2)
the adjoint code integrates time-dependent, non-linear and dissipative 
(hence physically irreversible) PDEs with
an explicit time integration loop running for \textit{ca.} $10^{6}$ time steps.
The approach relies on using the Adjoinable MPI library to
reverse the non-blocking communication patterns in the original code, and by controlling the 
memory overhead induced by the time-stepping loop with binomial checkpointing.
A description of the necessary code modifications is provided along with the
validation of the computed derivatives and a performance comparison of the tangent and adjoint codes.
\end{abstract}
\begin{keywords}
Algorithmic differentiation, computational fluid dynamics, sensitivity analysis, adjoint methods, 
parallel computing
\end{keywords}
\section{Introduction}
Numerical codes in engineering and physical sciences are most often used
to approximate the solution of governing equations on discretized domains.
A natural step beyond obtaining solutions in specific conditions is to seek
those conditions that modify the solution towards a specific goal, either
for optimization or control purposes. 
In this context, gradient-based methods play an important role. 
They require invariably the computation of derivatives, a task that can be automated by Algorithmic
Differentiation (AD) tools. In short, AD augments a given ``primal'' code that initially computes outputs
$Y_{i}$ from inputs $X_{j}$ into a ``differentiated'' code that additionally computes some derivatives
$dY_{i}/dX_{j}$ requested by the user. AD provides two main modes, the tangent/direct/forward mode and
the adjoint/reverse/backward mode.
If $1\leq i \leq p$ and $1\leq j \leq q$ bound the 
dimension of the output and input spaces, respectively, then the tangent mode is most efficient 
when $q\ll p$ while the adjoint mode is the only realistic option for $p\ll q$.
\par Many real life applications require computing derivatives of relatively
few outputs (cost functions, constraints\dots) with respect to many
inputs (state variables, design parameters, mesh coordinates\dots).
Adjoint AD would fit those applications perfectly since $p \ll q$. However,
two features of high-performance codes used in industry and academia
have been playing against that: parallel communications and unsteady
computations. The reason is the amount of resources needed to reverse
the data-flow and control-flow of such long and complex 
computations~\cite{griewank2008evaluating}.
As long as the adjoint mode provided by AD tools did not address
these serious limitations, most studies \cite{stampsAIAA2018,kenway2019effective,XU2015175}
circumvented them by using the following strategies:
\begin{itemize}
\item Applying AD on selected parts of the code without MPI calls and manually assembling the differentiated routines
  to obtain a correct adjoint code.
\item Restricting the use of AD to code solving problems that are either stationary or forced to become stationary.
  As an exception, earth sciences have long used adjoints of
  unsteady simulations \cite{Charpentier2001,HEIMBACH20051356}, pioneering the so-called checkpointing 
  schemes \cite{Gri92,GPR96_II} that we advocate here.
\item Using lower accuracy models
or otherwise ``compressed'' representations of the forward trajectory -- see, for example, \cite{GoetschelWeiser2015}. This approach has been frequently used as part of ``multifidelity'' modeling.
\end{itemize}
\par In this paper, we report on the outcome of exploiting the recently acquired maturity of
AD tools \cite{Huckelheim_multiactivity} at differentiating parallel code in adjoint mode 
by automatic inversion of the MPI communication layer \cite{Utke2009} 
while handling the cost of an unsteady computation.
\par The paper is organized as follows. In section~\ref{sec:background} we describe
the code to be differentiated and the current developments of adjoint AD for
parallel codes and unsteady computations.
In section~\ref{sec:test_cases} we introduce the two test cases analyzed,
section~\ref{sec:diff_workflow} explains in detail the process of code differentiation
that we followed, section~\ref{sec:results} presents our results and we offer our conclusions
in section~\ref{sec:conclusions}.
\section{Background} \label{sec:background}
\subsection{Primal code description}
\par The code under consideration in the present study - hereafter the \textit{primal}
code - is a computational fluid
dynamics (CFD) solver that integrates the governing equations for compressible fluid
flow. It belongs to a recent trend of CFD codes that use a
high-order spatial discretization adapted to 
compressible/incompressible flow computations on complex geometries modeled by unstructured meshes. 
This makes them 
suitable candidates to become industrial tools in the near future \cite{future_CFD}. 
Our application code is JAGUAR \cite{Jaguar,brunet2018comparison}, a solver for aerodynamics applications developed
to suit the future needs of the aerospace industry. 
Its excellent scalability, its ability to handle structured or unstructured grids,
its optimized 6-step time integration scheme as well as its high-order spatial
discretization based on spectral differences 
\cite{liu2006spectral,kopriva1998staggered,jameson2010proof,van2008stability,
vanharen2017revisiting} are all positive features 
which inevitably come at a price: the code is complex. Its manual 
differentiation for sensitivity computations is impractical, thus making AD
an attractive solution. The code is written following features from the Fortran 90
standard onward, with a zero-halo partitioning scheme and an MPI-based parallelisation. 
The version available for the present study used no thread pinning or dynamic 
load balancing.

\subsection{Overloading versus source-transformation AD tools}
\par AD can be based on two working principles: operator overloading (OO)
or source transformation (ST). 
The OO approach barely modifies the primal code: the data-type of numeric variables is
simply modified to contain their derivative in addition to their primal value, while
arithmetic operations are overloaded to act on both components of the variables.
While the debate is still active, it is generally agreed that ST AD tools require a
much heavier development, which is in general paid back by a better efficiency of the
differentiated code mostly in terms of memory consumption. Benchmark tests have pointed
out a tendency for OO-differentiated codes to be more memory demanding and somewhat
slower than their ST counterparts \cite{kenway2019effective}. 
%
%On the other hand, the higher flexibility of the OO model makes it almost
%readily applicable to languages with sophisticated constructs,
%such as C++ or Python, for which no ST tool exists to date.
%
On the other hand, the more flexible OO model can
be applied at a low
development cost to languages with sophisticated constructs,
such as C++ or Python, for which no general-purpose ST tool exists to date.
For a given application,
the choice between AD tools based on ST or OO is dictated by these constraints: with
JAGUAR being written in Fortran, ST appears to be the natural choice. Moreover,
for the size and number of time steps of our targeted applications, it is essential to
master the memory footprint of the final adjoint code. For this study, we have selected
the ST-based AD tool Tapenade \cite{Tapenade}. 
\subsection{AD of very long time-stepping sequences}\label{secAdjointTimeStepping}
\par The adjoint mode of AD leads to a code which executes the differentiated instructions in the reverse
order of the primal code. However, these differentiated instructions (the ``backward sweep'') use
partial derivatives based on the values of the variables from the primal code.
The primal code, or something close to it, must therefore be executed beforehand, forming the
``forward sweep''. As codes generally overwrite variables, a mechanism is needed to recover values
overwritten during the forward sweep, as they are needed during the backward sweep. Recovering
intermediate values can be done either by recomputing them at need, from some stored state, or by
storing them on a stack during the forward sweep and retrieving them during the backward sweep.
Neither option scales well on large codes, either with a memory use that grows linearly with the primal
code run time, or with an execution time that grows quadratically with the primal code run time. We
envision applications with 10$^5$ to 10$^6$ time steps to integrate the fluid flow equations. The
classical answer to this problem is a memory-recomputation trade off known as ``checkpointing'' \cite{griewank2008evaluating}.
A well-chosen checkpointing strategy can lead to execution time and memory consumption of the adjoint
code that grow only logarithmically with the primal code run time.
\par A checkpointing strategy is constrained by the structure of the primal code.
Checkpointing amounts to designating (nested) portions of the code, for which we are ready to pay
replicate execution to gain memory storage of its intermediate computations. These portions must have a
single entry point and a single exit point, for instance procedure calls or code parts that could be written
as procedures. For this reason one cannot in practice implement the theoretical optimal checkpointing scheme,
which is defined only on a fixed-length linear sequence of elementary operations of similar cost and nature.
A checkpointing scheme on a real code can still achieve a logarithmic behavior, but in general below the theoretical optimal.
Since checkpointing relies on repeated
execution, it requires storing and restoring a ``state'', which is a
subset of the memory and of the machine state such that the repeated
execution matches the original. This implies that the checkpointed code
portion is ``reentrant'', \textit{i.e.} avoids side-effects, so that 
running the portion twice does not alter the rest of the execution.
%
% Moreover, since checkpointed portions are supposed to be executed twice or more, they must be ``reentrant'':
% it must be possible to re-create the exact machine state at their entry point, and running them twice must not alter the
% rest of the execution. 
As a consequence, a checkpointed portion of code must always contain both
ends of an MPI communication, and similarly both halves of a non-blocking MPI communication~\cite{HascoetUtke2015}.
\par Time-stepping simulations are more fortunate: at the granularity of time steps, the code is indeed a fixed-length
sequence of elementary operations of similar cost and nature. The binomial checkpointing scheme \cite{walther2004advantages}
exactly implements
the optimal strategy in that case, and Tapenade applies it when requested.
Binomial checkpointing also reduces the state
restoration runtime overhead, through multiple restorations of each
stored state.
A checkpointing strategy is also constrained by the characteristics of the storage system. 
The binomial strategy assumes
a uniform and negligible cost for storing and retrieving the memory state before checkpoints (``snapshots''). This is in
reality never the case. New research \cite{aupy2016optimal,Aupy2017}
 looks for checkpointing strategies that take this memory cost into account,
as well as different access times for different memory levels. 
\par In general, few studies
confront unsteady problems directly, and most works reported in the literature focus
on problems around a fixed-point solution. Convergence towards that fixed steady state
is often enforced by means of implicit iterative schemes with preconditioning. Few iterations
are necessary, and only the final converged state requires storage before computing the
inverted set of instructions. Consequently, memory and computational overhead are kept low.
It is unfortunate, however, that many problems of industrial relevance are inherently unsteady.
In acoustics and combustion, for instance, unsteadiness simply cannot be ignored, which 
is what motivates the present study.
\subsection{Na\"ive \textit{vs.} selective AD} \label{sec:naive_vs_selective}
\par The AD tool requires that the function to be differentiated is designated in the primal code as
a ``head'' procedure. If the function is scattered over several code parts, one must modify
the primal code to make the head procedure appear -- an acceptable technical constraint.
The AD tool will then consider the complete sub-graph of the primal code's
call graph that is recursively accessible from the head. It will analyze it and differentiate
it as a whole, in one single step. Each procedure under the head will be differentiated with
respect to its ``activity pattern'', i.e. those of its inputs and outputs that are involved
in a dependence between the selected head procedure inputs and the selected head procedure outputs.
This approach is referred to as ``brute force'' \cite{JONES2011282} and ``full-code'' AD 
\cite{kenway2019effective} in two works that deem 
the approach inefficient, compared to manual assembly of individually differentiated procedures
with user-provided activity patterns. The inefficiency may come in part from the unavoidable
imprecision of the static data-flow analysis that detects the differentiated inputs and outputs of
each procedure. Static analysis has to make conservative approximations, that lead to larger
activity patterns than actually needed. One answer to that could be to use differentiation pragmas,
not provided at present in Tapenade. The inefficiency may also have come from the superposition of
activity patterns coming from a procedure's different call sites. The situation has now improved,
with Tapenade allowing a procedure to have several differentiated versions, potentially one
for each activity pattern encountered \cite{Huckelheim_multiactivity}.
\subsection{Parallel communications}
\par Parallelism is in general a challenge for AD, especially in adjoint mode, 
since variable reads give birth to adjoint variable increments. By design, the MPI distributed memory model
at least evacuates the risk of race conditions that would arise in other
models such as OpenMP.
% An additional challenge arises when the code to be differentiated by the AD tool contains message-passing
% instructions. 
Much conceptual work has been devoted to AD of MPI code \cite{HovlandBischof,
HeimbachHillGiering,schanen2010interpretative,schanen2012wish,schanen2010second}.
However, it is with the recent advent of the Adjoinable MPI library \cite{Utke2009} that 
several AD tools (Adol-C, Rapsodia, dco, OpenAD, Tapenade) support AD of code containing MPI
calls. Related projects include the Adjoint MPI library
\cite{towara2015mpi,schanen2010interpretative,schanen2010second} compatible with the
dco suite of AD tools, and CoDiPack \cite{sagebaum2017high} for C++ code - both based on OO. 
The automatic inversion of MPI calls necessary
to derive a parallel adjoint code can thus be performed by three tools. In \cite{towara2015mpi}, 
the CFD code OpenFOAM was adjointed with the combination of dco/c++ and Adjoint MPI.
CoDiPack was used to adjoint the CFD code SU2 \cite{sagebaum1}.
To our knowledge, these are the most similar studies to ours in terms of letting the AD tool
handle the parallel communication layer automatically - yet without
solving a time-dependent problem. 
\par 
An attractive choice is to restrict AD to parts of the code devoid of MPI communications.
The individually differentiated fragments can then be manually assembled
%Avoiding MPI idioms in code fed to an AD tool is an attractive choice.
%Individually differentiated routines can be manually assembled 
into an adjoint code that
preserves the often heavily optimized parallel communications layer of the primal code. 
A disadvantage of this approach is the increased workload incurred every time a different
problem is tackled, where the optimization concerns different quantities from those
previously considered. A certain degree of freedom in choosing the cost function and
automation in assembling differentiated procedures has been achieved in \cite{stampsAIAA2018}.
Alternatively, \cite{huckelheim_oms2018} has presented the transposed forward-mode
algorithmic differentiation to take advantage of those code portions featuring
symmetric properties in order to obtain adjointed code using the forward-mode AD.
Either way, handling MPI calls differently from the rest of the code contradicts the ultimate
goal of AD, which is full automation of the differentiation process regardless of the programming
features actually used in the primal code. It is true, however, that each specific library that
involves side-effects raises new issues, limitations, and challenges that cannot be readily
solved by AD tools. Given the efforts that have been devoted towards making MPI calls
compatible with AD tools~\cite{HovlandBischof,
HeimbachHillGiering,schanen2010interpretative,schanen2012wish,schanen2010second}, we aim
to test and document the outcome of letting the AD tool handle them alone. 
In this respect, our work intends
to provide a proof of concept illustrating that the route followed, which on the whole
has been avoided in the literature, is in fact practicable.
\section{Test cases} \label{sec:test_cases}
\subsection{Incompressible, viscous, two-dimensional double shear layer in a periodic square}
\par
\begin{figure}[h!]
  \includegraphics[trim = 0.2cm 0cm 0cm 0cm, scale = .85]{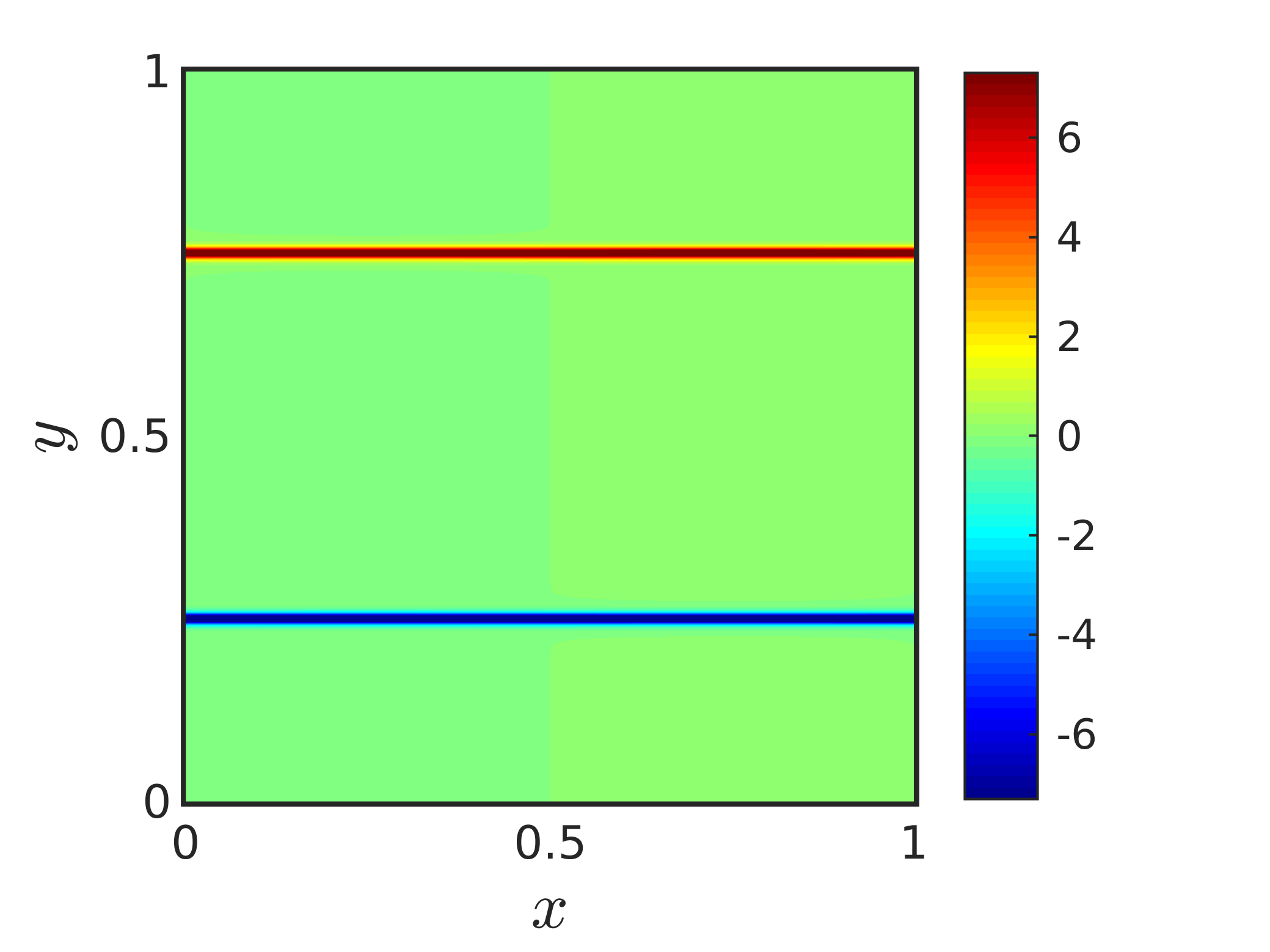}
  \includegraphics[trim = 1.2cm 0cm 0cm 0cm, scale = .85]{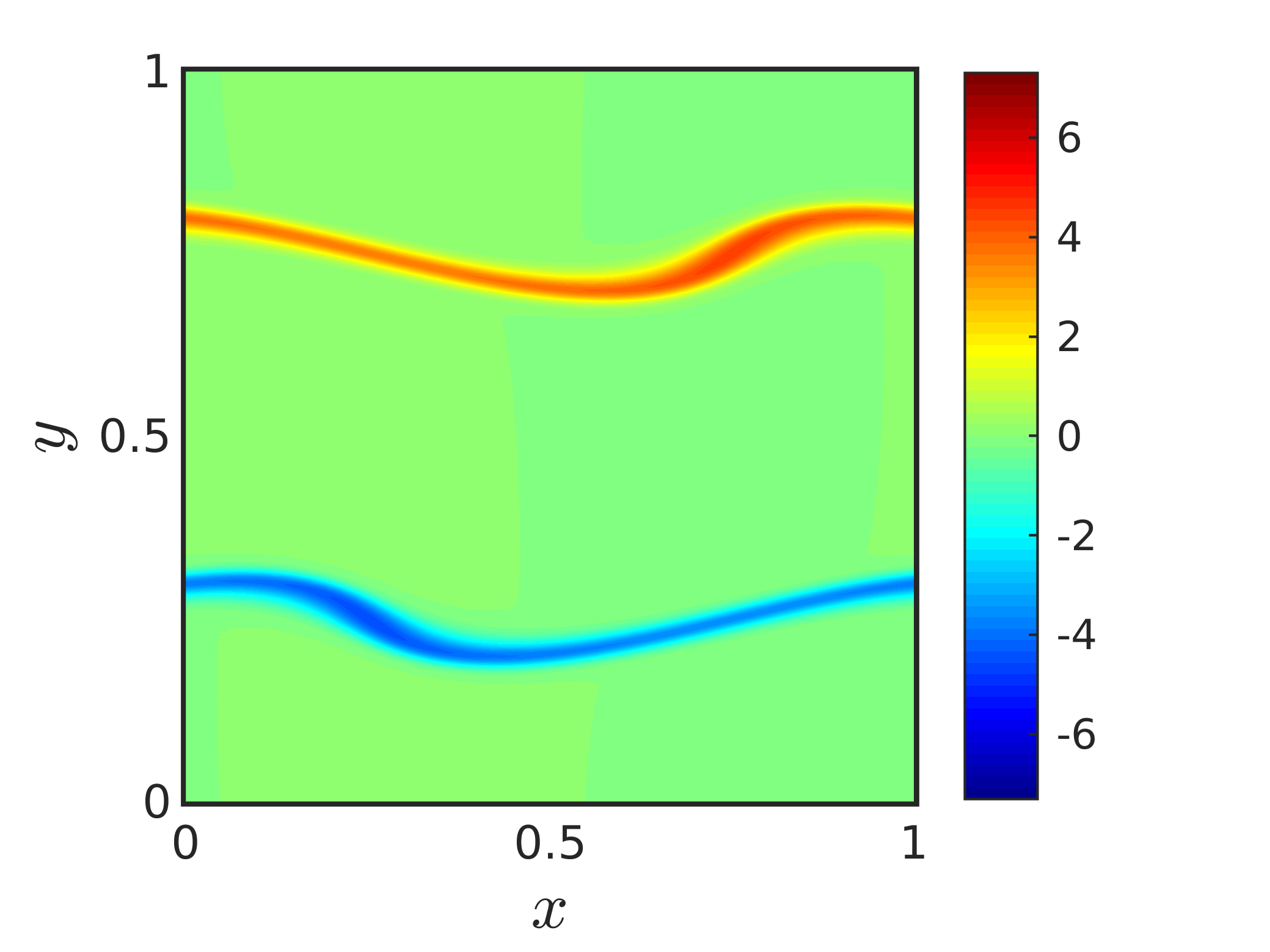}\\
  \includegraphics[trim = 0.2cm 0cm 0cm 0cm, scale = .85]{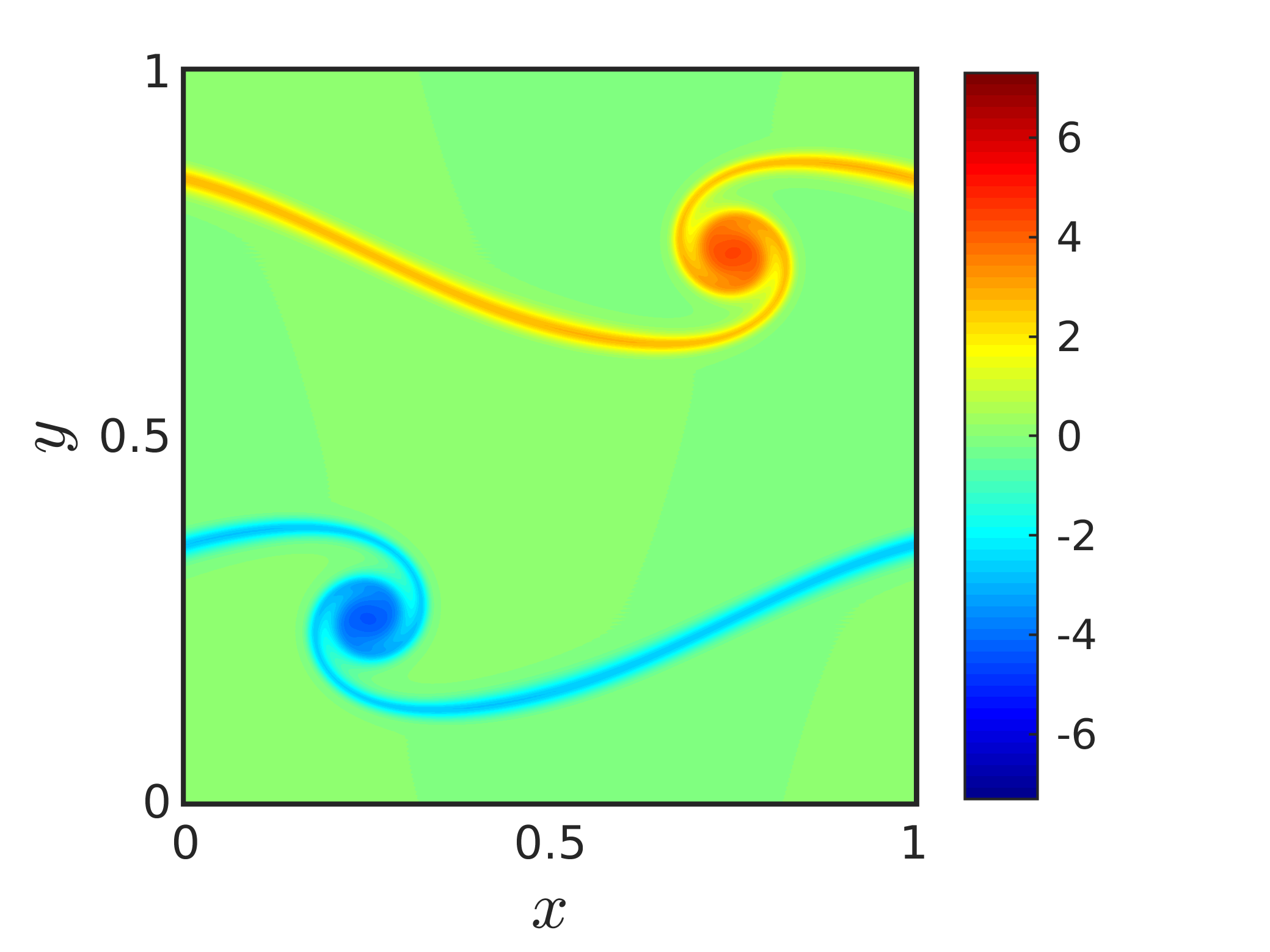}
  \includegraphics[trim = 1.2cm 0cm 0cm 0cm, scale = .85]{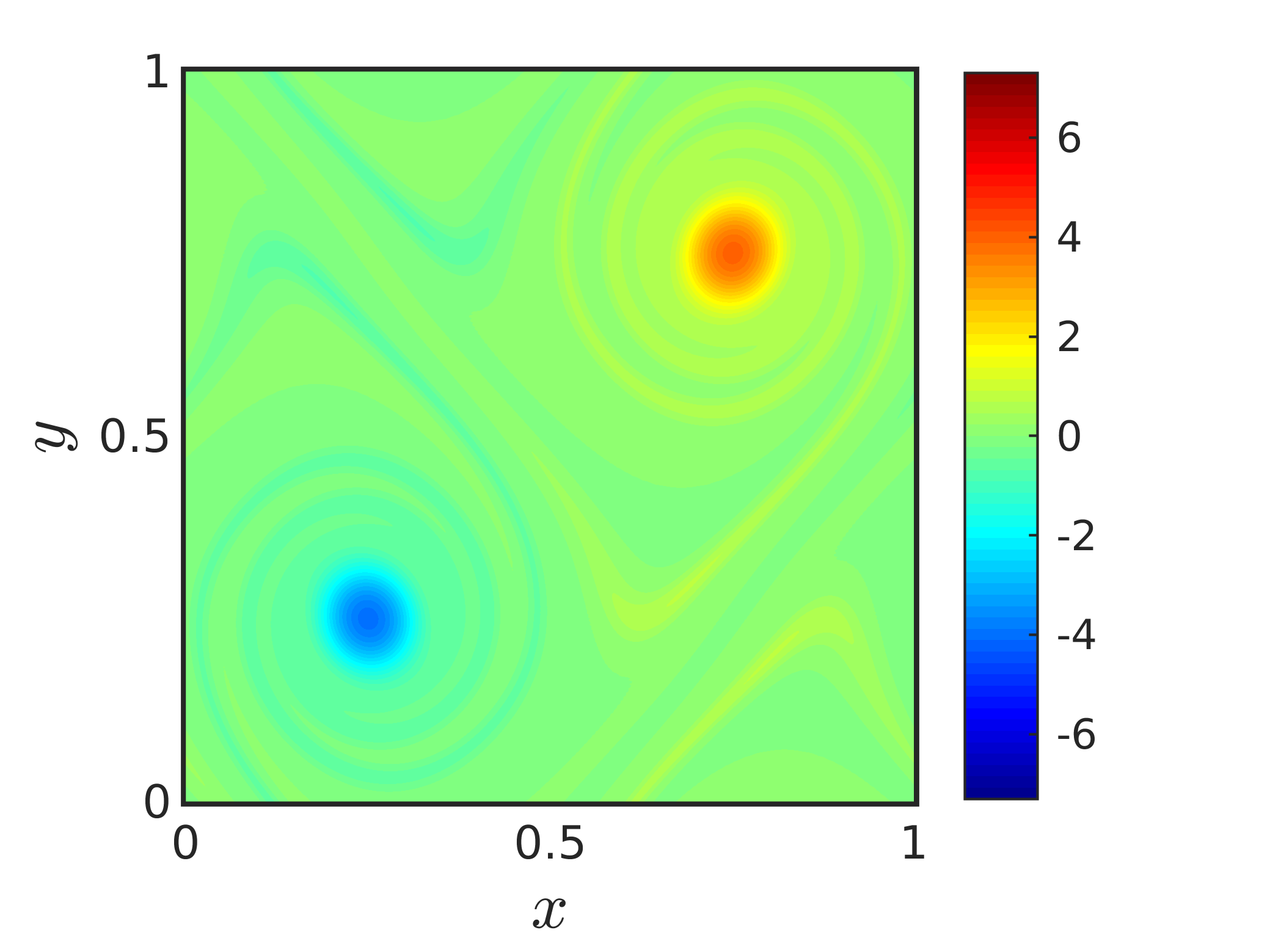}
  \caption{Spatial distribution of $\omega_{z}/\sqrt{\Omega_{ref}}$ for case $r160$ 
    at 4 instants $t\sqrt{\Omega_{ref}} = \{0,7,10,23\}$ in the following respective order: 
    top left, top right, bottom left and bottom right. 
    \label{fig:pcolors}}
\end{figure}

\begin{figure}[h!]
  \begin{center}
    \includegraphics[trim = 0.cm 0cm 0cm 0cm, scale = .5]{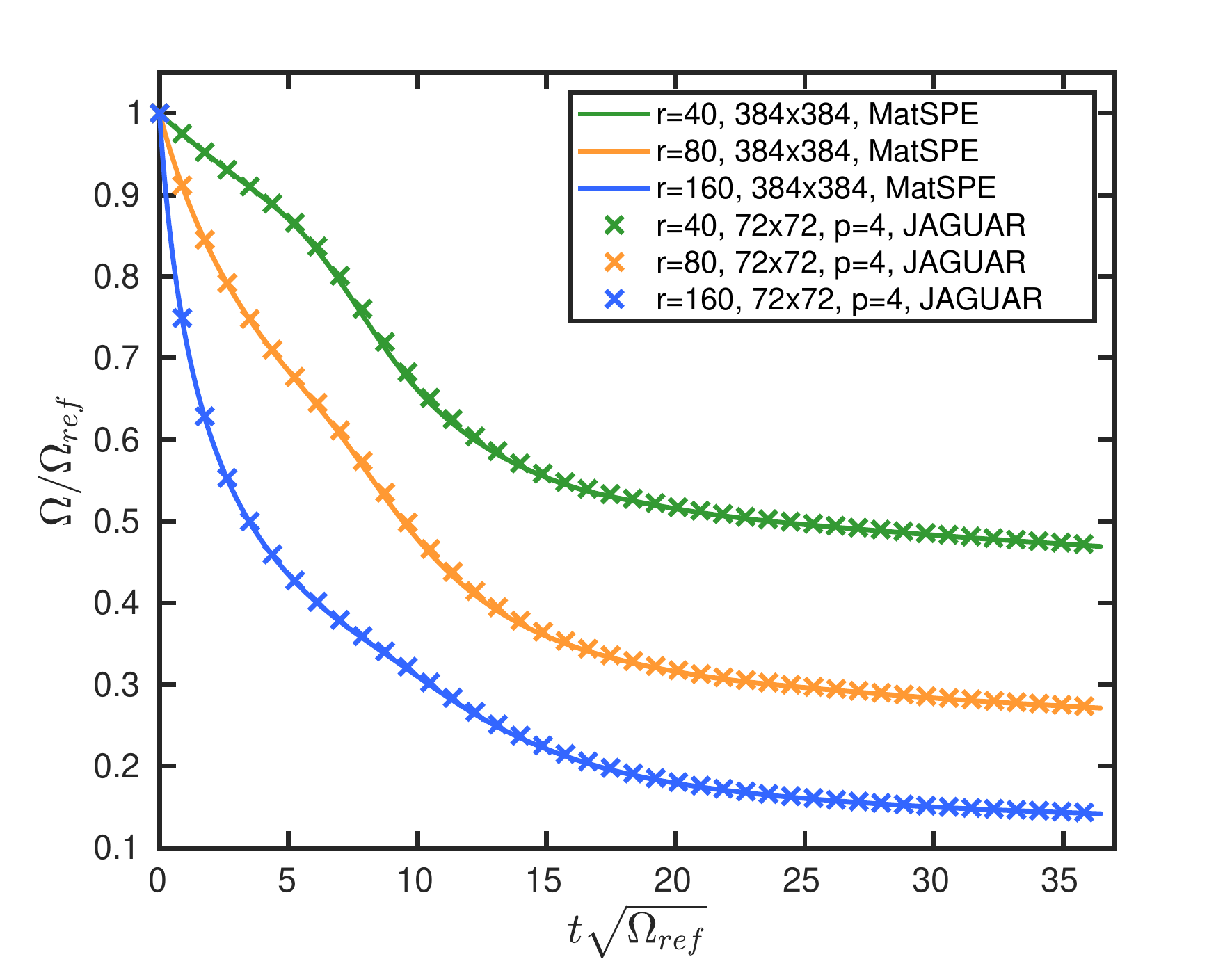}
  \end{center}
  \caption{Evolution of $\Omega(t)$ at 3 different values of $r$,
    JAGUAR \textit{vs.} fully spectral and incompressible code MatSPE. The legend includes
    the number of Fourier modes used in each direction for the MatSPE simulation (of which
    1/3 are zero-padded for dealiasing), while the JAGUAR data includes parameter $p$
    which is the selected order of the spatial discretization of the spectral difference
    scheme. The grid used in the JAGUAR simulation is a $72\times72$ structured mesh which,
    together with the setting $p=4$, yields $360$ degrees of freedom (DoF) per spatial direction. 
    So we are effectively comparing $256^{2}$ DoF with MatSPE against $360^{2}$ DoF with JAGUAR.\label{fig:sensis_long}}
\end{figure}

 We consider a viscous, two-dimensional incompressible flow in a square periodic
domain spanning $L=1$ in the streamwise $(x)$ and vertical $(y)$ directions. 
The velocity field at the initial instant $t_{0}$ is given by  
\begin{eqnarray}
  u &=& U\tanh\left[r\left(y-1/4\right)\right], \quad y\leq 1/2 \label{eqn:u_low}\\ 
  u &=& U\tanh\left[r\left(3/4-y\right)\right], \quad y>1/2 \label{eqn:u_up}\\ 
  v &=& U\delta\sin\left[2\pi\left(x+1/4\right) \right], \label{eqn:v}
\end{eqnarray}
where all quantities are made non-dimensional with $L$ and the streamwise reference
velocity $U_{0}=1$ as follows:
\begin{equation}
  \quad t=\tilde{t} \ U_{0}/L , \quad y=\tilde{y}/L, \quad x=\tilde{x}/L, 
  \quad U=\tilde{U}/U_{0}, \quad r=\tilde{r} \ L. 
\end{equation}
The parameters of the problem are $U$, $r$ and $\delta$. These are the
streamwise velocity amplitude, the shear parameter and the ratio of vertical to
streamwise velocity amplitudes, respectively. We set $\delta=0.05$ for the 
remainder of this study so that it is no longer a free parameter.
We analyze the evolution of the overall enstrophy $\Omega$, defined as 
\begin{equation}
\Omega = \int_{0}^{1}\int_{0}^{1}\frac{1}{2}\omega_{z}^{2} \ dx \ dy, \label{eqn:omega_def}
\end{equation}
where $\omega_{z}=\partial_{x}v-\partial_{y}u$ is the vorticity.
It can be readily shown from \Eqnrefs{eqn:u_low}-(\ref{eqn:v}) and (\ref{eqn:omega_def}) 
that at $t=t_{0}$,
\begin{equation}
  \Omega = U^{2}\left[ 6r\tanh(r/4)-2r\tanh^{3}(r/4) + 3\delta^{2}\pi^{2}\right]/3,
\label{eqn:omega_ref}
\end{equation}
and we choose $r=40$ with $U=1$ to yield an initial enstrophy level $\Omega_{ref}=53.36$
which we set as a constraint for all $r$. This implies $U$ is a function of $r$ only,
determined by re-arranging 
\Eqnref{eqn:omega_ref} as follows:
\begin{equation}
  U(r)= \left[ 3\Omega_{ref} \ / \ \left( 6r\tanh(r/4) -2r\tanh^{3}(r/4) + 3\delta^{2}\pi^{2}\right) \right]^{1/2}. \label{eqn:U_of_r}
\end{equation} 
The Reynolds number $Re_{0}=U_{0}L/\nu=1.176\times10^{4}$ is the same for all values of $r$
we consider, which are displayed on \Tabref{tab:settings}. 
\begin{table}[]
  \begin{center}
    \begin{tabular}{ c | c  c  c}
       case name & \textit{'r40'} & \textit{'r80'} & \textit{'r160'} \\
      \hline
      $r$ & 40 & 80 & 160 \\
      $U$ & 1 & 0.7072 & 0.5001\\
    \end{tabular}
    \caption{The three perturbation amplitudes $U$ corresponding to the 3 shear parameters $r$ 
      used in \Figref{fig:sensis_long}. $U$ is computed with \Eqnref{eqn:U_of_r}. 
      In all cases, the constraint of an initial enstrophy of $\Omega_{ref}=53.36$
      is applied.\label{tab:settings} }
  \end{center}
\end{table}
We display $\omega_{z}(x,y)$ for the case $r=160$ at four different instants on 
\Figref{fig:pcolors}, and $\Omega(t)$ for the three cases on
\Figref{fig:sensis_long}.
The time dependence of the flow is clear, albeit to a lesser extent in the final
stage where two large vortices slowly decay under the action of
viscosity. The motivation behind choosing this specific test case is that we
have effectively created an unsteady physical system where we can tune the dynamics with a single
parameter. We also intend to illustrate that a viscous flow, hence
dissipative and irreversible from a physical point of view, can safely be
treated by the adjoint-mode of AD executing the temporal integration in backwards mode.
It is sometimes pointed out that this type of problems can lead to unstable schemes
by drawing analogies with physical irreversibility and negative diffusivity.
Finally, we aim to show that unsteady problems governed by non-linear equations
do not necessarily lead to the failure of sensitivity analysis due to chaos, thus 
requiring shadowing techniques such as those in \cite{QiQiWang}. 
\par The incompressible 2D Navier-Stokes equations are solved with the initial and boundary conditions outlined
above using JAGUAR on a structured mesh with $72\times72$ square elements. The Mach number ($Ma$)
is set to zero to approach, as much as possible, incompressibility. 
The solution and flux points are located following Gauss-Lobatto-Chebyshev 
and Legendre collocation, respectively, and the CFL is kept constant at 0.5. The fluxes at the cell faces
are computed with the Roe scheme.
The temporal integration is done with a six-stage, fourth-order
low-dissipation low-dispersion Runge-Kutta scheme optimized for the spectral difference code
using the procedure in \cite{BERLAND20061459}.
We use an in-house, fully spectral code (MatSPE) designed for periodic incompressible viscous flows to solve the 
same test case and validate JAGUAR.
The output of the two codes is compared on \Figref{fig:sensis_long}, showing that 
the agreement between the codes is extremely good. 
\par We define the following cost function
\begin{eqnarray}
J &=& \int_{0}^{T}\Omega(t) \ dt. \label{eqn:cost_func1}
\end{eqnarray} 
Its derivative with respect to $r$ will be the target sensitivity
we compute by means of AD. From a physical point of view, $\Omega(t)$
is directly proportional to the rate of kinetic energy dissipation
in the flow due to the action of viscosity. 
Hence, the area under a curve of $\Omega(t)$ on \Figref{fig:sensis_long} for a given time interval
is a proxy for the kinetic energy dissipated by the flow during that time. 
It may be argued that since our numerical experiments only target $dJ/dr$,
it is pointless to use adjoint-mode AD because $r$ is a scalar. One must bear in mind, however,
that the final application of our study is to compute sensitivities of some $J$ with respect to
many inputs. Restricting the number of inputs to one in the sequel is only
a convenient way to validate our proposed work flow and the AD derivatives, as well as
to compare and study performance.
\par The computation of $J$ in the primal code is carried out by adding a contribution to the
time integral at each new time step in a running sum fashion, using a simple trapezoidal rule.
%
%Such a framework is convenient to illustrate the fundamental issue of adjoint-mode AD that we already
%described in section~\ref{secAdjointTimeStepping}. 
%
The tangent-differentiated code accumulates
contributions of each time step to $dJ/dr$, along with the primal time-stepping sequence, \textit{i.e.}
in the same order. This is conceptually simple: once we reach the iteration corresponding to $t = T$
(which we will call iteration number $N$), 
both $J$ and $dJ/dr$
are known and the program can end. In contrast, the adjoint-differentiated code will first run an
initial forward sweep that will integrate the Navier-Stokes equations from $t=0$
to iteration $N$ of the time-stepping loop,
chiefly to generate the final state of the program variables. Only then can the backward
sweep of the adjoint code start to accumulate derivatives, computing the sensitivity of $J$ with
respect to the state variables at iteration $N-1$, and carry on stepping back in time to finally
obtain $dJ/dr$ when $t=0$ is reached.
In order to provide intermediate values from the forward sweep to the backward
sweep in the correct order (\textit{i.e.} reversed), a combination of stack storage and additional forward
recomputation is needed, making a good checkpointing scheme essential. The recomputations and stack use
will inevitably imply that one
run of the adjoint code requires significantly more memory and execution time than
the tangent code. We thus expect the tangent code to still outperform the adjoint code when $p=q$ or when $q$
is only a few times larger than $p$, but the adjoint code will definitely outperform the tangent code when
$p \ll q$, which is the case in many applications.
\subsection{Inviscid compressible flow over an airfoil section}
\par
\begin{figure}[h!]
  \begin{center}
    \includegraphics[trim = 4cm 0cm 0cm 0cm, scale = 0.25]{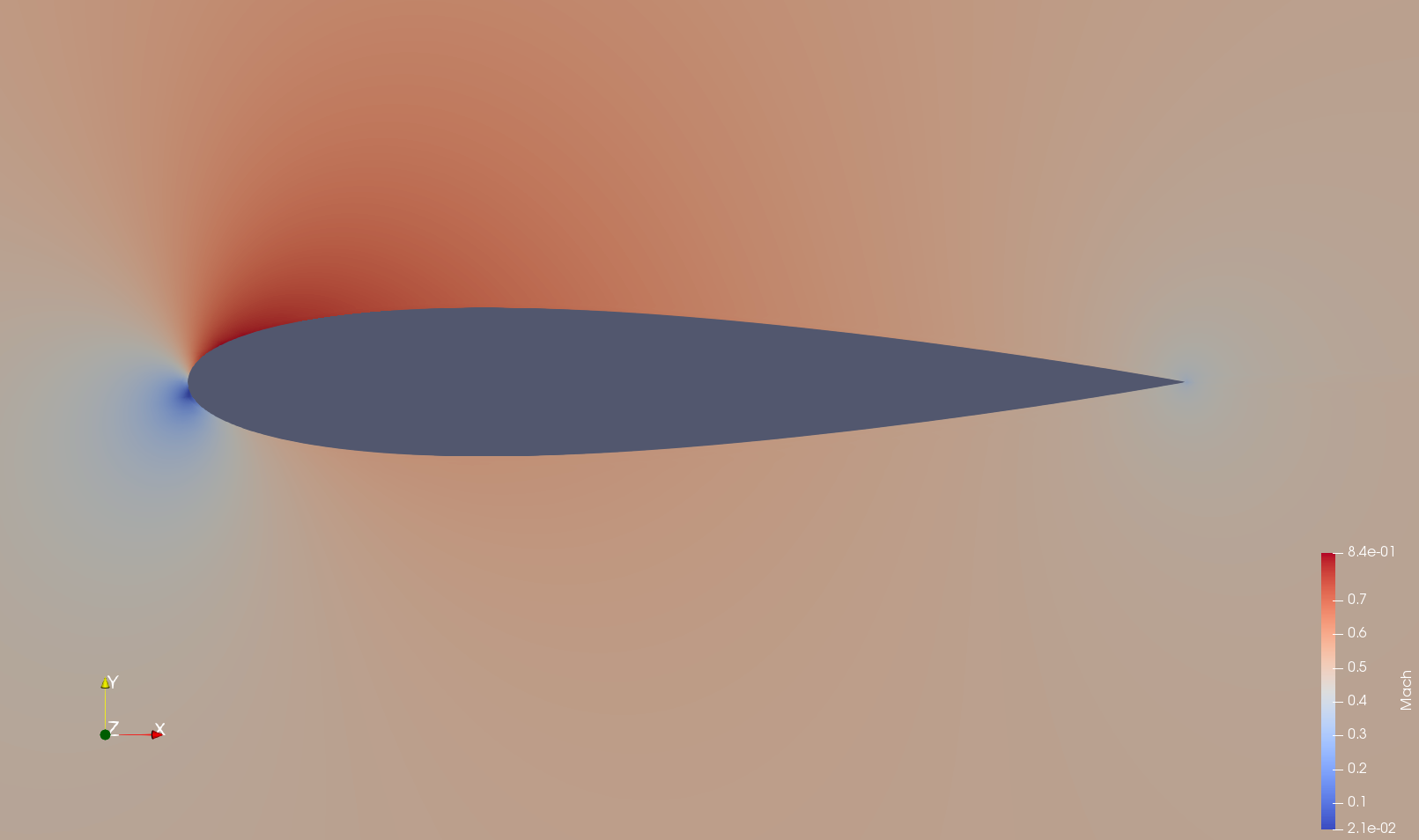}
  \end{center}
  \caption{Distribution of the Mach number around a NACA 0012 airfoil section at $\alpha=4^{\circ}$,
    computed with JAGUAR in inviscid two-dimensional mode. The upstream Mach number is $Ma=0.5$, the flow
    is from left to right.
    \label{fig:NACA_Mach_pcolor}}
\end{figure}

 We study the two-dimensional flow around a NACA 0012 airfoil at zero angle of attack $\alpha$.
It is a symmetric airfoil commonly used as an aerodynamics test case.
Even though the flow eventually
becomes stationary, JAGUAR is run in unsteady mode so that the steady state is
reached without numerical stabilization. Two important differences with respect to the previous
test case deserve to be highlighted: the flow is compressible  
and there is no viscosity in the set of equations solved by JAGUAR. 
We consider two levels of compressibility: close to incompressible with $Ma=0.1$,
and a higher subsonic regime with $Ma=0.5$. Both are below the critical Mach number, so 
no shock appears on the upper side of the airfoil. Yet at $Ma=0.5$, we expect to see
sizable departures from predictions based on two-dimensional, incompressible and inviscid (\textit{i.e.}
potential) flow predictions.
The domain is discretized with
a structured mesh of 2876 quadrilateral elements. We are interested in the derivative
of the lift coefficient $Cl$ with respect to $\alpha$ at
$\alpha=0^{\circ}$. This frequently
computed quantity allows for comparison with experimental measurements
\cite{mccroskey1987technical} and with numerical computations based on the
potential flow solver XFOIL \cite{XFOIL}. \Figref{fig:NACA_Mach_pcolor}
shows the Mach number distribution around the airfoil for the angle of attack
$\alpha=4^{\circ}$, computed by JAGUAR with the same mesh as the $\alpha=0^{\circ}$ case.
At this $\alpha$ setting, an asymmetric pressure distribution gives rise to lift
due to the faster flow on the upper side -- yet subsonic conditions are mantained. 
\subsection{Sensitivity validations with finite differences}
\par The estimates of a given sensitivity computed with tangent- or adjoint-mode AD
codes should agree almost to machine precision, as they result from the same computation
modulo associativity-commutativity. In contrast, the reference value against which to validate the
sensitivity is obviously a finite-difference (FD) estimate, which is subject to errors
due to the contribution of higher derivatives. For the viscous test case, we compute our FD estimates with two independent
realizations at $r$ and $r+dr$, where $dr/r = 10^{-5}$. Similarly, 
for the inviscid test case we use two independent
realizations at $\alpha$ and $\alpha+d\alpha$, where $d\alpha/\alpha = 10^{-5}$.
We thus expect an agreement
between FD and AD derivatives up to more or less half of the decimals, whereas we expect a much better
agreement between tangent AD and adjoint AD derivatives.
%
% \pagebreak
\section{Differentiation work flow} \label{sec:diff_workflow}
\par We adopted the following work flow with regards to the JAGUAR flow solver, 
the working principles of Tapenade and the Adjoinable MPI library:
\begin{itemize}
\item[A)] Identify the part of the code that computes the function to
differentiate, exactly from the differentiation input variables to the
differentiation output, and make it appear as a procedure (the ``head''
procedure of section~\ref{sec:naive_vs_selective}). This may require a
bit of code refactoring. The differentiation tool must be given (at least)
this root procedure and the call tree below it.
\item[B)] Migrate all MPI calls to Adjoinable MPI, whether AD will be applied in
tangent or in adjoint mode. This involves two steps.
First, as Adjoinable MPI does not support all MPI communication styles
(e.g. one-sided), the code must be transformed to only use the supported
styles, which is a reasonably large subset. Second, effectively
translate the MPI constructs into their Adjoinable MPI equivalent, which
occasionally requires minor modifications to the call arguments. As
Adjoinable MPI is just a wrapper around MPI, the resulting code should
still compile and run, and it is wise to test that.
\item[C)] Provide the AD tool with the source of the head procedure and of all
the procedures that it may recursively call, together with the
specification of the differentiation input and output variables. Then differentiate, 
after which two steps follow. First, fix all issues signaled by the AD tool, e.g.
unknown external procedures or additional info needed, and validate the
differentiated code. Second, address performance issues and in
particular optimize the checkpointing strategy by adding AD-related
directives to the source. This may also involve special treatment of
linear algebra procedures such as solvers.
\end{itemize}
\par Step A) is best illustrated with the viscous test case. 
The main program in JAGUAR calls thirteen procedures before $r$
is used to initialize the velocity field. $r$ has thus no impact on the code prior to the 
velocity field initialization. A few procedure calls after the initialization, the time
integration routine is called, at the end of which $J$ is known. All code after that point
has no impact on the sensitivity computation we investigate.
So we create a head that contains all procedure calls from the velocity field
initialization to the time integration routine, excluding everything else.
This head, say \texttt{TopSub(X,Y)},
takes $X=r$ as input and outputs $Y=J$. It will be differentiated in step C).
\par The work involved in step B) is likely to depend heavily on the primal code and the state
of advancement in the Adjoinable MPI project at the time of implementation. 
Codes tuned for high-performance computing which repeat a non-blocking communication pattern
many times benefit from using persistent communication requests through the
\texttt{MPI\_SEND\_INIT} and \texttt{MPI\_SEND\_RECV} calls.
This leads to the later use of the \texttt{MPI\_STARTALL} and \texttt{MPI\_WAITALL} constructs
when the communications are actually invoked. These communication patterns are not currently
supported by the Adjoinable MPI library, and the advice of the developers in such cases is to
replace them by loops over processes containing individual non-blocking \texttt{MPI\_ISEND},
\texttt{MPI\_IRECV} and their respective \texttt{MPI\_WAIT} calls. The output of the
modified primal code needs to be validated against the original version, which can be done
reliably on a small number of processes. The impact of the modifications on the performance,
however, is difficult to assess without access to hundreds, preferably thousands of cores.
With 16 cores, we record a performance drop of $2.6\%$ after $3.5\times 10^{5}$ time steps
on the inviscid test case. Since the modified code preserves the non-blocking structure of the
original code, we expect the performance drop to be low. The drop should be attributed to 
the overhead for communication between the process and the communication controller, which is
what persistent communications alleviate - and not between one communication controller and another.
All in all, support for persistent communications would be a useful addition to the Adjoinable MPI
library.
\par Migrating from MPI to Adjoinable MPI involves changes that are mainly cosmetic. The name of each MPI
call requires
relabeling to match the name of the Adjoinable MPI wrapper for that call. Typically, all that needs
to be done is to change \texttt{CALL MPI\_COMM\_SIZE}, \texttt{CALL MPI\_BARRIER},
\texttt{CALL MPI\_ALLREDUCE}, \textit{etc.} into \texttt{CALL AMPI\_COMM\_SIZE},
\texttt{CALL AMPI\_BARRIER}, \texttt{CALL AMPI\_ALLREDUCE} and so on. These modifications
can be automated with a suitable script.
For each point-to-point Adjoinable MPI call, however, the user has to manually add an 
argument that specifies the kind of MPI call at the other end of the communication. 
For example, the extra argument for a send instruction may be \texttt{AMPI\_TO\_RECV}, 
\texttt{AMPI\_TO\_IRECV\_WAIT} or \texttt{AMPI\_TO\_IRECV\_WAITALL}.
Similarly, the extra argument for a receive may be \texttt{AMPI\_FROM\_SEND}, 
\texttt{AMPI\_FROM\_ISEND\_WAIT}, \texttt{AMPI\_FROM\_ISEND\_WAITALL}. This must
be left to the programmer, as no static data-flow analysis can determine it in general.
% For MPI ``sends'' and ``receives'', however, the user has to
% manually add an additional argument to the Adjoinable MPI call, which depends on the
% communication pattern involved. They are documented in the FAQ section of the Tapenade
% website\footnote{https://www-sop.inria.fr/tropics/tapenade.html}, and they must be left to the programmer
% because it is in general impossible to identify the communication pattern by static data-flow analysis.
%
\par In step C), specific
instructions can be passed on to tune how Tapenade should handle the checkpointing of the
primal code. The most important part in specifying these
is to identify the time-stepping loop(s), to label them for binomial
checkpointing, and to assign them a well-chosen number of snapshots.
It can be done by placing a number of specific directives
in the source code, to indicate which subroutine calls should be checkpointed portions. By default,
all of them are. This is a delicate trade off, of experimental nature. The rule of thumb is to avoid
checkpointing on ``small'' procedures, as they require very little space on stack and on the other
hand may require a large snapshot for checkpointing.
Typically a subroutine that takes in an array and an index to just overwrite the
array element at this index is considered ``small'' and should not be
checkpointed.
A case of forbidden checkpointing, detailed in \cite{HascoetUtke2015}, is a subroutine
that contains one end of a non-blocking communication, for example the
{\tt MPI\_ISEND} or the {\tt MPI\_IRECV}, but not the corresponding {\tt
MPI\_WAIT}. The subroutine is not reentrant and must not be checkpointed.
In our application, this happened on two procedures. 
%
% The adjoint-mode differentiation requires switching off checkpointing
% of two procedures for correct execution when non-blocking communication calls in the primal
% code are split between different procedures. For instance, when the \texttt{MPI\_ISEND} and \texttt{MPI\_IRECV}
% are in one subroutine while their corresponding \texttt{MPI\_WAIT} is inside
% another subroutine. This scenario was flagged in \cite{HascoetUtke2015} as problematic, hence
% the necessary workaround. 
%
For completeness, we must point out a manual modification required on the adjoint code, i.e. after
step C).
At the ``turning'' point between forward and backward sweeps, one must insert specific calls to declare,
for each primal variable {\tt V} that is active (\textit{i.e.} it has a derivative) and that is passed through MPI,
the correspondence with its derivative counterpart {\tt Vb}. Specifically, this is done by calling
{\tt ADTOOL\_AMPI\_Turn(V, Vb)}. This must be done by hand, until Tapenade generates these calls automatically.
\par The differentiation of the head routine \texttt{TopSub(X,Y)} leads to 
\texttt{TopSub\_d(X,Xd,Y,Yd)} in tangent mode and to \texttt{TopSub\_b(X,Xb,Y,Yb)} in adjoint mode. 
The final program calls the differentiated routines when derivatives are needed. In doing so,
it is necessary to correctly set the initial \texttt{Xd} (respectively \texttt{Yb}) and to correctly
interpret the resulting \texttt{Yd} (resp. \texttt{Xb}). For parallel
execution specifically, the full sensitivity could be spread across the
\texttt{Xb} of several processes
after having seeded all of them with \texttt{Xb=0} and \texttt{Yb=1}.
In such cases, a global reduction operation
is needed to sum the resulting \texttt{Xb} held by each process in order to
recover the correct sensitivity.
%
% \par The differentiation of the head \texttt{topsub(X,Y)} will lead to \texttt{topsub\_d(X,dX,Y,dY)} in tangent mode and
% \texttt{topsub\_b(X,dX,Y,dY)} in adjoint mode. These differentiated versions of the head
% need to be embedded inside a \textit{driver} routine that calls them after setting the seed values
% for $dX$ and $dY$. The driver should,
% of course, contain the procedure calls of the primal code which were not differentiated.
% An effect to consider is that in adjoint mode, the sensitivity is not entirely contained inside
% variable $dX$ as one would expect when differentiating a serial code. Instead, after having seeded
% all processes with $dX=0$ and $dY=1$, a reduction operation is necessary to sum the $dX$ held
% by each process in order to recover the correct sensitivity. 
%
Strategies for handling the
dispersion of the derivatives across different processes in adjoint-mode AD have been discussed
in \cite{schanen2016semantics}.
\par Let us finally comment on our experience using an AD tool on a
large parallel code. 
%Since two authors are CFD specialists and
%one is an AD tool researcher and developer, we cannot claim objectivity.
% However, we can state that the difficulties encountered on the AD tool
% side, including 3 bug reports, will no longer stand on the way of future
% attempts at performing a similar study.
The difficulties encountered on the AD tool
side, including 3 bug reports, will no longer stand on the way of future
attempts at performing a similar study.
This gives a rough idea of
the maturity, or lack thereof, of the Tapenade tool. 
Issues with the Adjoinable MPI library were limited to
the lack of support, at the time of writing, for persistent communication 
requests. 
The fact that Tapenade
does not yet introduce automatically the needed calls to {\tt ADTOOL\_AMPI\_Turn}
is a limitation. We would like to think that having one developer of
the AD tool among us is not a prerequisite for success. The three bugs
mentioned above were fixed after being reported by email. 
Performance tuning of the adjoint code is not yet optimal,
and we believe this calls for better support from the AD tool for
choosing a good checkpointing scheme.
\section{Results} \label{sec:results}
\subsection{Viscous double shear layer}
\begin{table}[h!]
  \begin{center}
    \begin{tabular}{l | l }
       Differentiation method & Sensitivity $(dJ/dr)$\\
      \hline
      FD (MPI)          & -0.22002\textbf{254}  \\
      Tangent AD (AMPI) & -0.22002394265\textbf{381}    \\
      Adjoint AD (AMPI) & -0.22002394265\textbf{861}    \\ 
    \end{tabular}
    \caption{$dJ/dr$ computed with three different methods, for the time
      integration interval between $t=0$ and $t=T$ ($6.8\times 10^{5}$ time steps). Viscous test case with $r=160$, run on 16 parallel processes. 
      \label{tab:goal2_results} }
  \end{center}
\end{table}
\par The temporal integration of the equations of motion
is carried out from $t=0$ to the $n-th$ iteration of the time integration
loop in JAGUAR, with $n=6.8\times10^{5}$. With the CFL setting outlined in
section \ref{sec:test_cases}, this number of time steps corresponds to
$t\sqrt{\Omega_{ref}}=19.8$, which from \Figref{fig:pcolors} can be seen
as the time when the decay of $\Omega$ becomes slow for all three values of $r$.
The value of $dJ/dr$ is given in
Table~\ref{tab:goal2_results}, where the results from FD, tangent-mode AD
and adjoint-mode AD are all gathered. 
The agreement between FD-based sensitivities and AD validates the differentiation procedure.
In particular, we note that although the underlying system is unsteady 
and results from non-linear equations, 
there is no breakdown of the conventional sensitivity
analysis. It is important to emphasize this because, as illustrated by \cite{QiQiWang}
for the Lorentz attractor, unsteadyness and chaos can undermine sensitivity computations
and motivate the use of more complex approaches. But our system, although unsteady and
non-linear, does not fall into the category of problems that require these approaches. 
\par The agreement between the two AD-based sensitivities is excellent, within round-off error
of double precision arithmetic. It was expected in case of correct differentiation
by Tapenade, but nevertheless it is puzzling when comparing the drastic
differences between the two differentiated codes.
The fact that the adjoint-differentiated code outputs the correct answer after carrying out
the time-stepping loop backwards confirms the absence of any stability or convergence issues
related to inverting the instructions of a code that integrates in time an irreversible and 
dissipative system. More specifically, there is no issue of numerical instability caused by a term with negative 
diffusivity.
\par The time required for the various computations is shown on Table~\ref{tab:goal2_comptimes}.
A factor of two is indicated for the FD computation, given that two runs of the primal
code are required. 
The computations are run on 16 Intel(R) Xeon(R) Gold 6140 processors at 2.30GHz.
The Intel Fortran compiler version 18.0.2 is used, with identical optimization flags
for all codes: \texttt{-ipo -O3}. The Intel MPI Library was used for the MPI implementation, in its 2018 version.
It appears that the tangent-mode AD can be faster than two runs of
the primal code, requiring 1.7 times the execution time of the primal code. 
The higher accuracy of sensitivity computations from tangent-mode AD thus comes with 
the added benefit of a faster computation than that of FD. Furthermore, each
additional cost function differentiated with respect to $r$ will require
an additional FD computation, whereas the cost of each new sensitivity with
respect to $r$ will keep the cost of the tangent-mode execution constant.
We note in passing that the FD computation is based on the MPI code before
the modifications of step B) outlined in section~\ref{sec:diff_workflow},
so that the tangent-mode AD is faster than the FD
computations {\em despite} the move from MPI to Adjoinable MPI.
This indicates that the cost of this additional wrapper on top of the MPI library is negligible.
%
%
% FD: jaguar-masterFinnMPI_3/formation/Viscous/shear2D_jic/verb_680000_np16_r160.0016.txt, 38121s-38202s, jfonc =    52.1554310248266
% FD: jaguar-masterFinnMPI_3/formation/Viscous/shear2D_jic/verb_680000_np16_r160.txt 37304s-37441s, jfonc = 52.1557830605420
%     Time take: (37304+38202)/2 = 37753s
%     sensi = (52.1554310248266-52.1557830605420)/0.0016=-0.220022322126034
%
%     
%
% TGT:jaguar-masterFinnMPI_7/formation/Viscous/shear2D_jic/verb_680000steps_np16_bis.txt, 71646s-71753s, jfoncd = -0.220023942656334
% TGT:jaguar-masterFinnMPI_7/formation/Viscous/shear2D_jic/verb_680000steps_np16.txt, 71796s-72035s, jfoncd = -0.220023942656334
%     Time take: (71646+72035)/2 = 71840s
%
%
%
% ADJ:jaguar-masterFinnMPI_9/formation/Viscous/shear2D_jic/verb_680000_snap20_np16.txt, 629164s-630003s, sum slwb =  -0.220023942653187
%     Time take: (629164+630003)/2 = 629583s
%
%
\begin{table}[h!]
  \begin{center}    
    \begin{tabular}{l | c }
       Differentiation method & Normalized compute time \\
      \hline
      FD (MPI) & $1\: (\times 2)$  \\
      Tangent AD (AMPI) & $1.7$  \\
      Adjoint AD (AMPI) & $15.4$        
    \end{tabular}
    \caption{Same as \Tabref{tab:goal2_results}, but showing execution times normalized
      by the primal code execution time. Viscous test case, run on 16 parallel processes.
      \label{tab:goal2_comptimes} }
  \end{center}
\end{table}
\par Table~\ref{tab:goal2_comptimes} also shows the slowdown factor of the adjoint code, which is
15.4 and deserves some discussion. An initial experiment without any
specific checkpointing scheme simply ran out of memory after only less
than a hundred time-stepping iterations. Therefore, binomial
checkpointing is unavoidable. It accounts for a significant part of this
adjoint slowdown: since we chose to allow for 80 snapshots for binomial
reversal of $6.8\times 10^{5}$ time steps, the binomial model tells us this costs
an average 3.9 extra recomputations per time step. This leaves us with
roughly an 11--fold slowdown to account for, which is still higher than
expected. By far the most expensive task in JAGUAR is the computation of the right-hand
side of the governing equations, required six times per time step by
calling procedure \texttt{ComputeRhsNavier}. The latter can itself be decomposed into eight
subroutines, shown on \Figref{fig:computerhs}.
\begin{figure}[h!]
  \begin{lstlisting}
    SUBROUTINE ComputeRhsNavier
       CALL ExtrapolSolAndComputeFlux
       CALL Scatter
       CALL RiemannSolver
       CALL GatherSolution
       CALL GradientAndInternalFlux
       CALL ViscousFlux
       CALL GatherFlux
       CALL FluxDivAndUpdate
    END SUBROUTINE ComputeRhsNavier
  \end{lstlisting}
  \caption{Breakdown of the most computationally-expensive subroutine in the code,
    \texttt{ComputeRhsNavier}.
    \label{fig:computerhs}}
\end{figure}
Upon differentiation in adjoint mode, the
subroutines within \texttt{ComputeRhsNavier} will lead to adjoint versions, such as
\texttt{ExtrapolSolAndComputeFlux\_b}, \texttt{Scatter\_b}, \textit{etc}.
\Tabref{tab:adjoint_slowdowns} contains the slowdown factor for each individual 
adjoint subroutine found inside \texttt{ComputeRhsNavier} compared to its original version.
\begin{table}[]
  \begin{center}
    \begin{tabular}{ l | c}
        \hspace{1.3cm} subroutine & adjoint/primal slowdown  \\
      \hline
       \texttt{ExtrapolSolAndComputeFlux} & 2.35\\
       \texttt{Scatter} & 1.69\\
       \texttt{RiemannSolver} & 8.45\\
       \texttt{GatherSolution} & 2.56\\
       \texttt{GradientAndInternalFlux} & 12.84 \\
       \texttt{ViscousFlux} & 6.44\\
       \texttt{GatherFlux} & 1.73 \\
       \texttt{FluxDivAndUpdate} & 5.60\\
    \end{tabular}
    \caption{Slowdown factor between original and adjoint versions of the subroutines
      within \texttt{ComputeRhsNavier}. None of the routines above include
      parallel communication calls within them. \label{tab:adjoint_slowdowns} }
  \end{center}
\end{table}
$54\%$ of the time taken by \texttt{ComputeRhsNavier} is spent
during execution of \texttt{GradientAndInternalFlux}, so that the largest 
slowdown -- almost 13 -- occurs precisely within the most expensive routine.
Calls to push and pop intermediate values into the AD stack are spread across 
all eight procedures, and for this reason we cannot ascribe the performance
penalty incurred specifically in \texttt{GradientAndInternalFlux} to AD stack access
only. 
However, we do note that improving the performance when reading and writing into
the AD stack -- which happens regardless of binomial checkpointing -- will generally 
have a direct impact on adjoint-differentiated code speed.
Future efforts on the AD tool side will be geared in that direction.
\par The type of information on \Tabref{tab:adjoint_slowdowns} provides the AD user
with indications on where to hand-tune the AD process further -- by preventing some
procedures from being checkpointed, for instance.
But from here onwards, the process is no longer automatic and will depend
heavily on the specific code under consideration. 
We therefore limit our study to the performance obtained
``out of the box'' with the current degree of maturity in the development of the AD
tool, which has undergone major improvements over the years to reach its present
state.
With the obtained performance, the adjoint code is
already preferable to the tangent code as soon as the number of input
variables, with respect to which we request
sensitivities, goes over 15.
\subsection{Inviscid compressible flow around an airfoil}
\begin{figure}[h!]
  \begin{center}
    \includegraphics[trim = 0.cm 0cm 0cm 0cm, scale = 0.8]{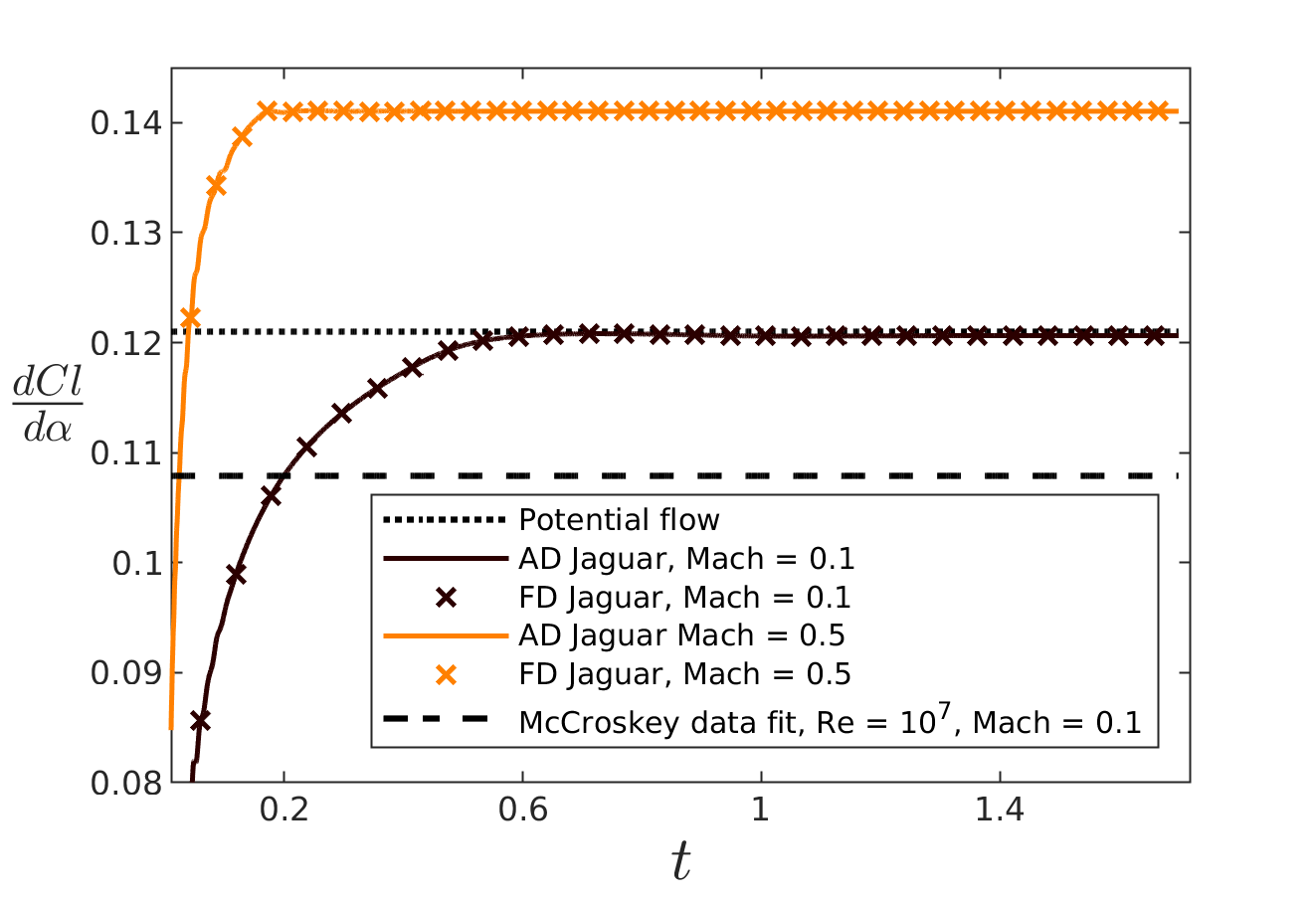}
  \end{center}
  \caption{Sensitivity of the lift coefficient $Cl$ to a change in incidence angle $\alpha$ 
    around a NACA 0012 airfoil at zero incidence. Two Mach numbers considered, sensitivities
    computed with FD and JAGUAR differentiated in tangent-mode AD. Values are compared
    with the curve fit to experimental data from \cite{mccroskey1987technical}, and with
    the solution found using FD with the panel code XFOIL (potential flow). Time $t$ 
    is in code units. We verified that the agreement between the JAGUAR output at $Ma=0.1$ 
    and $Ma=0.5$ is considerably improved if normalized by the low $Ma$ correction
    factor $\sqrt{1-(Ma)^{2}}$.
    \label{fig:NACA_0012sensis}}
\end{figure}
\par The flow around the airfoil converges to a steady state after integrating the
governing equations for long enough. It can be seen on Figure~\ref{fig:NACA_0012sensis} 
that both the FD approximation and the AD computation (only tangent-mode used) 
exhibit similar time-dependence during an initial transient phase, and then stabilize
around a converged value. The FD and AD computations agree across both Mach numbers and
throughout the time interval considered, validating the AD code. We also show the value 
obtained for the sensitivity using an altogether different code (XFOIL) and applying
FD to it around the same airfoil geometry and for an angle of attack $\alpha=0$. 
The experimental curve fit
shown on Figure~\ref{fig:NACA_0012sensis} is for comparison, illustrating that the
discrepancy between the various estimations
of the same quantity is within reasonable bounds considering the difference between
the approaches. 
\par The time span on Figure~\ref{fig:NACA_0012sensis} by the JAGUAR computations
represents $4.3\times10^{5}$ and $1.2\times10^{6}$ time steps with Mach=0.1 and 0.5,
respectively, using a CFL of 0.5 in both cases. 
The tangent-mode AD computations last 1.6 times the execution time of the corresponding
primal code for both Mach numbers. For this test case, the computations are run on
27 Intel® Xeon(R) Silver 4110 processors at 2.10GHz,
with the Intel Fortran compiler version 15.0.1 and identical optimization flag
for primal and AD codes: \texttt{-O3}. MPICH-3.2 was used. It is interesting to see
that the ratio of run times between the tangent-differentiated and the primal code is very close to
that obtained for the previous test case, which was 1.7. The primal codes of both test cases 
differ due to the presence of additional code lines related to the viscous stresses. 
\section{Conclusions and future work} \label{sec:conclusions}
\par A CFD code with an optimized parallel communications layer has been 
automatically differentiated
by letting the AD tool handle the communications layer in an automated way. 
% A CFD code with a high-order spatial discretization based on spectral differences
% and an optimized parallel communications layer has been automatically differentiated
% by letting the AD tool handle the communications layer in an automated way. 
In adjoint
mode the inversion of the communications during the backward sweep was found to produce
correct code which could execute in 15 times the primal code
execution time. The computational overhead is to a large extent the consequence of having
to resort to binomial checkpointing to trade storage for computational time in order to
invert a temporal integration loop with a number of iterations of the order of $10^{6}$.
We have presented a detailed outline of the code modifications required to achieve the
correct differentiation of the parallel code. Two flows were solved with the CFD code,
an inviscid and a viscous test case. The latter is a physically dissipative system,
which has been computed using the backward mode of AD without running
into stability issues and yielding the correct derivative at the end of the computation.
Both test cases exhibited run times for the tangent-differentiated codes which were 
in the range 1.6-1.7 times slower than a single primal code run. They are therefore readily
superior to finite difference approximations even for single derivative computations.  
A natural further step would involve embedding the derivative solver into an optimal
control loop. The strength of a code such as JAGUAR lies in its ability
to handle acoustics problems, such as the noise radiated by the wake
of an object during a given time. Optimization in this type of time-dependent and 
multi-parameter applications is the subject of ongoing work.
\section{Acknowledgments}
This work has been financed by a grant from the STAE Foundation for the
3C2T project, managed by the IRT Saint-Exup{\'e}ry. J.I.C. acknowledges funding
from the People Program (Marie Curie Actions) of the European Union's 
Seventh Framework Program (FP7/2007-2013) under REA grant agreement 
n. PCOFUND-GA-2013-609102, through the PRESTIGE program coordinated by 
Campus France. The computations for the viscous test case were carried out at
the CALMIP computing center under the project P0824.

\bibliography{mybibfile}
\bibliographystyle{plain}

\end{document}